\documentclass[11pt,a4paper]{article}
\pdfoutput=1

\usepackage[nosort]{cite}
\usepackage[bulletsep]{collref}

\usepackage[british]{babel} %dates in british format

\usepackage{epsfig,rotating}
\usepackage{amsmath,amssymb,amsthm}
\usepackage{amsfonts}
\usepackage{mathrsfs}
\usepackage{bbm}
\usepackage{bm}

\usepackage{color}
\usepackage[textwidth = 430 pt, textheight = 630 pt]{geometry}
\definecolor{MyDarkBlue}{rgb}{0.15,0.25,0.45}

\usepackage[linktocpage=true]{hyperref}
\hypersetup{
colorlinks=true,
citecolor=MyDarkBlue,
linkcolor=MyDarkBlue,
urlcolor=MyDarkBlue,
pdfauthor={Christian S\"amann, Martin Wolf},
pdftitle={Non-Abelian Tensor Multiplet Equations from Twistor Space},
pdfsubject={hep-th}
breaklinks=true
}

\let\fn\footnote
\renewcommand{\footnote}[1]{\linespread{1.1}\fn{#1}\linespread{1.29}}

\makeatletter\renewcommand{\section}{\@startsection
{section}{1}{\z@}{-3.5ex plus -1ex minus
    -.2ex}{2.3ex plus .2ex}{\bf\mathversion{bold} }}
\makeatletter\renewcommand{\subsection}{\@startsection{subsection}{2}{\z@}{-3.25ex
plus -1ex minus
   -.2ex}{1.5ex plus .2ex}{\bf\mathversion{bold} }}
\makeatletter\renewcommand{\subsubsection}{\@startsection{subsubsection}{3}{-2.45ex}{-3.25ex
plus -1ex minus -.2ex}{1.5ex plus .2ex}{\it }}
\renewcommand{\thesection}{\arabic{section}}
\renewcommand{\thesubsection}{\arabic{section}.\arabic{subsection}}
\renewcommand{\@seccntformat}[1]{\@nameuse{the#1}.~~}

\renewcommand{\theequation}{\thesection.\arabic{equation}}
\makeatletter \@addtoreset{equation}{section}

\renewcommand*\l@section{\@dottedtocline{1}{0em}{2em}}
\renewcommand*\l@subsection{\@dottedtocline{2}{2em}{2.4em}}
\renewcommand*\l@subsubsection{\@dottedtocline{4}{3.8em}{3.7em}}

\renewcommand\tableofcontents{%
    \section*{\large\contentsname
        \@mkboth{%
          \MakeUppercase\contentsname}{\MakeUppercase\contentsname}}%
       {\baselineskip=15pt plus 2pt minus 1pt
    \@starttoc{toc}}%
}

\renewenvironment{thebibliography}[1]
     {\baselineskip=16pt plus 2pt minus 1pt
      \section*{\large\refname
        \@mkboth{\MakeUppercase\refname}{\MakeUppercase\refname}}%
     \list{\@biblabel{\@arabic\c@enumiv}}%
           {\settowidth\labelwidth{\@biblabel{#1}}%
            \leftmargin\labelwidth
            \advance\leftmargin\labelsep
            \@openbib@code
            \usecounter{enumiv}%
            \let\p@enumiv\@empty
            \renewcommand\theenumiv{\@arabic\c@enumiv}}%
      \sloppy
      \clubpenalty4000
      \@clubpenalty \clubpenalty
      \widowpenalty4000%
      \sfcode`\.\@m
 \catcode`\^^M=10%
}

\newcommand{\appendices}{
\section*{Appendix}\label{appendices}\setcounter{subsection}{0}
\addcontentsline{toc}{section}{Appendix}
\setcounter{equation}{0}
\makeatletter
\renewcommand{\theequation}{\Alph{subsection}.\arabic{equation}}
\renewcommand{\thesubsection}{\Alph{subsection}}
\@addtoreset{equation}{subsection}
\makeatother
}

\numberwithin{lemma}{section}

\numberwithin{definition}{section}

\newtheorem{theorem}{Theorem}
\numberwithin{theorem}{section}

\numberwithin{prop}{section}

\numberwithin{cor}{section}

\def\periodb#1{\setbox0=\hbox{$#1$}#1\hskip-\wd0\hbox to\wd0{-}}

\newcommand{\unit}{\mathbbm{1}}   			% identity map/matrix
\newcommand{\CF}{\mathcal{F}}

\newcommand{\CN}{\mathcal{N}}
\newcommand{\CO}{\mathcal{O}}

\newcommand{\frg}{\mathfrak{g}}				% frak-letters
\newcommand{\frh}{\mathfrak{h}}				% frak-letters

\newcommand{\FC}{\mathbbm{C}}     			% field of complex numbers
\newcommand{\PP}{{\mathbbm{P}}}    			% complex projective plane
\newcommand{\dd}{\mathrm{d}}     			% total differential
\newcommand{\dpar}{\partial}     			% partial differential
\newcommand{\diag}{{\mathrm{diag}}}     		% diagonal matrix
\newcommand{\eps}{{\varepsilon}}			% antisymmetric tensors
\newcommand{\ald}{{\dot{\alpha}}}     			% dotted letters
\newcommand{\bed}{{\dot{\beta}}}
\newcommand{\gad}{{\dot{\gamma}}}

\newcommand{\eand}{{~~~\mbox{and}~~~}}     		% and etc. in equations
\newcommand{\eon}{{~~\mbox{on}~~}}     		% and etc. in equations
\newcommand{\ewith}{{~~~\mbox{with}~~~}}

\newcommand{\der}[1]{\frac{\dpar}{\dpar #1}}   		% partielle ableitung, 1 argument
\newcommand{\au}{\mathfrak{u}}
\newcommand{\asu}{\mathfrak{su}}

\newcommand{\sU}{\mathsf{U}}     			% groups

\newcommand{\sSU}{\mathsf{SU}}
\newcommand{\sSL}{\mathsf{SL}}

\newcommand{\sAut}{\mathsf{Aut}}

\newcommand{\sG}{\mathsf{G}}
\newcommand{\sH}{\mathsf{H}}

\newcommand{\sB}{\mathsf{B}}

\newcommand{\sSp}{\mathsf{Sp}}
\newcommand{\sOSp}{\mathsf{OSp}}

\newcommand{\acton}{\vartriangleright}     			% span
\newcommand{\remark}[1]{}     				% remark
\def\tyng(#1){\hbox{\tiny$\yng(#1)$}}			% small Young diagram
\def\tyoung(#1){\hbox{\tiny$\young(#1)$}}			% small Young diagram
\newcommand{\mt}{\mathsf{t}}
\renewcommand{\sb}{\mathsf{b}}
\newcommand{\sft}{\mathsf{t}}

\begin{document}
\begin{titlepage}

\setcounter{page}{0}
\renewcommand{\thefootnote}{\fnsymbol{footnote}}

\begin{flushright}
 EMPG--12--09\\ HWM--12--06\\ DMUS--MP--12/05
\end{flushright}

\begin{center}

{\LARGE\textbf{\mathversion{bold}Non-Abelian Tensor Multiplet Equations\\ from Twistor Space}\par}

\vspace{1cm}

{\large
Christian S\"amann$^{a}$ and Martin Wolf$^{\,b}$
\footnote{{\it E-mail addresses:\/}
\href{mailto:c.saemann@hw.ac.uk}{\ttfamily c.saemann@hw.ac.uk}, \href{mailto:m.wolf@surrey.ac.uk}{\ttfamily m.wolf@surrey.ac.uk}
}}

\vspace{1cm}

{\it
$^a$ Maxwell Institute for Mathematical Sciences\\
Department of Mathematics,
Heriot--Watt University\\
Edinburgh EH14 4AS, United Kingdom\\[.5cm]

$^b$
Department of Mathematics,
University of Surrey\\
Guildford GU2 7XH, United Kingdom\\[.5cm]

}

\vspace{1cm}

{\bf Abstract}
\end{center}
\vspace{-.3cm}

\begin{quote}
We establish a Penrose--Ward transform yielding a bijection between holomorphic principal 2-bundles over a twistor space and non-Abelian self-dual tensor fields on six-dimensional flat space-time. Extending the twistor space to supertwistor space, we derive sets of manifestly $\CN=(1,0)$ and $\CN=(2,0)$ supersymmetric non-Abelian constraint equations containing the tensor multiplet. We also demonstrate how this construction leads to constraint equations for non-Abelian supersymmetric self-dual strings.
\vfill
\noindent 6th May 2014

\end{quote}

\setcounter{footnote}{0}\renewcommand{\thefootnote}{\arabic{thefootnote}}

\end{titlepage}

\tableofcontents

\bigskip
\bigskip
\hrule
\bigskip
\bigskip

\section{Introduction}

Gauge theories describe the dynamics of both a connection on a principal bundle and a set of fields that form sections of associated vector bundles. The connection defines parallel transport of fields along paths in space-time. Locally, a connection is given by a Lie algebra-valued differential 1-form called the connection 1-form. Taking the path-ordered exponential of the integral of this 1-form along the path, we obtain a group element describing the effect of the parallel transport (holonomy).

Consider now the generalisation of parallel transport from point-like objects, which we parallel transport along paths, to one-dimensional objects such as strings, which we parallel transport along surfaces. Analogously to the integral over the connection 1-form, we expect here to integrate over a surface and therefore the existence of a 2-form potential. Such a 2-form potential appears in the connective structure of a gerbe, and gerbes will thus replace the notion of principal bundles.

Abelian gerbes appear in various guises, e.g.\ in terms of stacks of groupoids \cite{0817647309} or the bundle gerbes of \cite{Murray:9407015}. The non-Abelian case is more intricate, as it requires the introduction of a `surface ordering'. This problem is overcome by the non-Abelian gerbes of \cite{Breen:math0106083,Aschieri:2003mw}. These non-Abelian gerbes are special cases of so-called principal 2-bundles \cite{Bartels:2004aa}, which are categorified bundles in the sense of \cite{Baez:2003,Baez:2003aa}.\footnote{See \cite{Nikolaus:2011} for a comprehensive account on the different approaches.} Theories describing the dynamics of the connective structure on principal 2-bundles\footnote{or, more generally, on principal $n$-bundles} are known as higher gauge theories \cite{Baez:2010ya}. Particularly interesting are higher gauge theories in six dimensions with a 2-form potential that has a self-dual 3-form field strength. In the following, we will refer to general theories containing a self-dual 3-form field strength as self-dual tensor field theories. Such theories are expected to play an analogous role in the description of M5-branes and NS5-branes in type IIA string theory as Yang--Mills theories do for D-branes. Various self-dual tensor field theories have been proposed recently, see e.g.\ \cite{Ho:2011ni,Samtleben:2011fj,Chu:2011fd,Chu:2012um,Samtleben:2012mi}, however, very few use the framework of higher gauge theory, e.g.\ \cite{Fiorenza:2012tb,Palmer:2012ya}.

As we shall demonstrate, a useful guiding principle for the development of non-Abelian self-dual tensor field theories is the twistor approach of \cite{Saemann:2011nb,Mason:2011nw} (see \cite{Baston:1989,Chatterjee:1998} for an earlier account). There it was shown that holomorphic Abelian 1-gerbes over a suitable twistor space are bijectively mapped to Abelian 2-form potentials with self-dual 3-form field strength via a Penrose--Ward transform. Replacing the Abelian gerbes by holomorphic principal 2-bundles, we therefore expect to be able to establish a bijection between such 2-bundles and solutions to the field equations of a self-dual tensor field theory based on higher gauge theory. In addition, if we replace the twistor space by the appropriate supertwistor space, we should obtain the supersymmetric extensions of these equations.

As a result directly derived from this twistor description, we should also be able to discuss non-Abelian supersymmetric self-dual strings. Recall that the self-dual string \cite{Howe:1997ue} is a field configuration of a 2-form potential with a self-dual 3-form field strength, which is translationally invariant along a temporal and a spatial direction. Such configurations are thus obtained by a dimensional reduction of a self-dual tensor field theory, and this dimensional reduction is easily implemented at the level of twistor geometry as shown in \cite{Saemann:2011nb}.

In this paper, we shall address the extensions of the twistor construction developed in \cite{Saemann:2011nb,Mason:2011nw} to both the non-Abelian case as well as the supersymmetric setting. We start in Section \ref{sec:review} with a brief review of the notion of principal 2-bundles and connective structures. We establish the Penrose--Ward transform for a (purely bosonic) non-Abelian tensor field equation in Section \ref{sec:PWbos}. In Section \ref{sec:PWSUSY}, we extend the Penrose--Ward transform supersymmetrically to the cases of both $\CN=(1,0)$ and $\CN=(2,0)$ supersymmetry and construct non-Abelian constraint equations on (chiral) superspace involving the tensor multiplet. We also discuss the reduction of our constructions to the self-dual string equation in Section \ref{sec:sds} before concluding in Section \ref{sec:conc}.

\section{Higher gauge theory with principal 2-bundles}\label{sec:review}

In the transition from gauge theory to higher gauge theory, we have to replace principal bundles by principal 2-bundles. The gauge groups, which are given by the structure groups of principal bundles, are correspondingly replaced by Lie crossed modules.\footnote{A Lie crossed module is equivalent to a strict Lie 2-group. This motivates our notation in this paper. Furthermore, we shall use the terms Lie crossed module and strict Lie 2-group interchangeably.} We first review Lie crossed modules and their linearisations before coming to the definition of principal 2-bundles. Our discussion will be very concise; for further details, see e.g.~\cite{Bartels:2004aa,Breen:2006,Wockel:2008aa,Baez:2010ya}.

\subsection{Lie crossed modules}

A {\em Lie crossed module} is a pair of Lie groups $(\sG,\sH)$ together with an automorphism action $\acton$ of $\sG$ on $\sH$ and a group homomorphism $\sft:\sH\rightarrow \sG$, where for all $g\in\sG$ and $h,h_1,h_2\in\sH$,
\begin{itemize}
\item[i)] $\mt$ is equivariant with respect to conjugation,
\begin{subequations}\label{eq:LCMIdentities}
\begin{equation}
 \mt(g\acton h)\ =\ g \mt(h) g^{-1}~,
\end{equation}
\item[ii)] and the so-called {\em Peiffer identity} holds:
\begin{equation}
 \mt(h_1)\acton h_2\ =\ h_1h_2h_1^{-1}~.
\end{equation}
\end{subequations}
\end{itemize}
In general, we will denote a Lie crossed module by $(\sH\overset{\sft}{\to}\sG,\acton)$ or simply by $\sH\to\sG$.
A simple example of a Lie crossed module is the {\em inner automorphism Lie 2-group} $C_{\sG,{\sf Inn}}:=(\sG\overset{\sft}{\to}\sG,\acton)$ where $\sG$ is some Lie group, $\sft$ is the identity and $\acton$ is the adjoint action. We can extend this to the {\em automorphism Lie 2-group} $C_{\sG,\sAut}:=(\sG\overset{\sft}{\to}\sAut(\sG),\acton)$ where $\sG$ is some Lie group, $\sAut(\sG)$ are its automorphisms, $\sft$ is the obvious embedding via the adjoint action and $\acton$ is the canonical action. Other interesting examples of Lie crossed modules include the central extension Lie 2-group $C_N:=(\sU(N)\overset{\sft}{\to}\sSU(N),\acton)$ with $\sft$ and $\acton$ being the obvious projection and the adjoint action, respectively, as well as the shifted version\footnote{Note that every Abelian Lie group $\sG$ is a Lie 2-group $\sB\sG$ over a one element set with both $\sft$ and $\acton$ being trivial. This shift is the first step in horizontal categorification.} of $\sU(1)$, $\sB\sU(1)=(\sU(1)\overset{\sft}{\to}\{\unit\},\acton)$, where both $\sft$ and $\acton$ are trivial. 

Linearising both groups in a Lie crossed module $(\sH\overset{\sft}{\to}\sG,\acton)$ at their identity elements, we obtain the notion of a differential crossed module\footnote{A differential crossed module is equivalent to a strict Lie 2-algebra or a 2-term $L_\infty$-algebra with vanishing Jacobiator. As in the finite case, we shall use the terms differential Lie crossed module and strict Lie 2-algebra interchangeably.} which is an $L_\infty$-algebra that plays the role of the gauge algebra. A {\em differential crossed module} $(\frh\overset{\sft}{\to}\frg,\acton)$ is a pair of Lie algebras $(\frg,\frh)$ with an action $\acton$ of elements of $\frg$ as derivations of $\frh$ and a Lie algebra homomorphism $\sft:\frh\rightarrow \frg$, which satisfy the linearised versions of equations \eqref{eq:LCMIdentities}. That is, we have
\begin{equation}
 \mt(X\acton Y)\ =\ [X,\mt(Y)]\eand \mt(Y_1)\acton Y_2\ =\ [Y_1,Y_2]
\end{equation}
for all $X\in \frg$ and $Y,Y_1,Y_2\in \frh$. The differential crossed module corresponding to $C_{\sG,\mathsf{Inn}}$ is $(\frh\overset{\sft}{\to}\frg,\acton)$, where $\frh=\frg=\mathsf{Lie}(\sG)$, $\sft$ is the identity, and $\acton$ the adjoint action. The linearisation of the Lie crossed modules $C_N$ and $\sB\sU(1)$ are $c_N:=(\au(N)\overset{\sft}{\to}\asu(N),\acton)$ and $\sb\au(1):=(\au(1)\overset{\sft}{\to}0,\acton)$ with the obvious maps $\sft$ and $\acton$. Interestingly, the 3-algebras that appear in M2-brane models are also special cases of differential crossed modules \cite{Palmer:2012ya}.

\subsection{Principal 2-bundles}

In the following, let $M$ be a manifold with an open (Stein) covering $\mathfrak{U}=\{U_a\}$ that we choose sufficiently fine in each situation. Recall that a principal bundle $E$ over $M$ with structure (Lie) group $\sG$ is a manifold that is locally diffeomorphic to the spaces $U_a\times \sG$. The spaces $U_a\times \sG$ are then patched together to the total space of $E$ by the transition functions $g_{ab}$. The transition functions are given by an element of the first non-Abelian \v{C}ech cohomology $H^1(M,\sG)$. That is, on non-empty overlaps $U_a\cap U_b$ we have smooth maps $g_{ab}:U_a\cap U_b\to\sG$  such that the following cocycle conditions are satisfied\footnote{Intersections of coordinate patches will always be assumed to be non-empty.}
\begin{equation}
 g_{ab}\ =\ g_{ba}^{-1}~~~\mbox{on}~~~U_a\cap U_b\eand g_{ab}g_{bc}g_{ca}\ =\ \unit~~~\mbox{on}~~~U_a\cap U_b\cap U_c~.
\end{equation}
Two cocycles $g_{ab}$ and $\tilde g_{ab}$ are considered equivalent (or cohomologous), if we have a set  of smooth maps $g_a:U_a\to \sG$ such that
\begin{equation}
  g_{ab}\ =\ g_a \tilde g_{ab} g_b^{-1}~.
\end{equation}
A {\em trivial principal bundle} has therefore transition functions $g_{ab}$ that can be split according to
\begin{equation}
 g_{ab}\ =\ g_a g_b^{-1}~.
\end{equation}
In this case, we shall also write $g_{ab}\sim\unit$.

Analogously, we define principal 2-bundles $E$ with strict structure Lie 2-groups $(\sH\overset{\sft}{\to}\sG,\acton)$ by transition functions $g_{ab}:U_a\cap U_b\to \sG$ and $h_{abc}:U_a\cap U_b\cap U_c\to \sH$. These represent elements of a generalised, non-Abelian  \v{C}ech cohomology $H^2(M,\sH\to\sG)$ that is defined by the following cocycle conditions:
\begin{equation}\label{eq:cocylce}
 \sft(h_{abc})g_{ab}g_{bc}\ =\ g_{ac}\eand h_{acd}h_{abc}\ =\ h_{abd}(g_{ab}\acton h_{bcd})
\end{equation}
on appropriate overlaps. Two sets of transition functions $(g_{ab},h_{abc})$ and $(\tilde g_{ab},\tilde h_{abc})$ are considered equivalent (or cohomologous), if there are smooth maps $g_a:U_a\to \sG$ and $h_{ab}:U_a\cap U_b\to \sH$ such that
\begin{equation}\label{eq:equivalence}
 g_a \tilde g_{ab}\ =\ \sft(h_{ab})g_{ab}g_b\eand h_{ac}h_{abc}\ =\ (g_a\acton \tilde h_{abc})h_{ab}(g_{ab}\acton h_{bc})~.
\end{equation}
Using these transformations, we can always set $h_{aaa}=\unit$, which, in turn, induces $g_{aa}=\unit$ and $h_{aab}=h_{abb}=\unit$. We shall always do so in the following. Residual transformations that remain are those with $h_{aa}=\unit$. In addition, a {\em trivial principal 2-bundle} has transition functions $(g_{ab},h_{abc})$ such that there are smooth maps $(g_a,h_{ab})$ satisfying
\begin{equation}\label{eqTriv2B}
 g_a\ =\ \sft(h_{ab})g_{ab}g_b\eand h_{ac}h_{abc}\ =\ h_{ab}(g_{ab}\acton h_{bc})~.
\end{equation}
Here, we have used that $g\acton\unit=\unit$.

Note that a principal 2-bundle with $C_{\sG,\sAut}$ as its strict structure Lie 2-group corresponds to the non-Abelian gerbes of \cite{Breen:math0106083} and \cite{Aschieri:2003mw} while Abelian gerbes (as the bundle gerbes of \cite{Murray:9407015}) are obtained from principal 2-bundles with the strict structure Lie 2-group $\sB\sU(1)$. Note also that analogously to the notion of holomorphic principal bundles, which come with holomorphic transition functions, we can introduce holomorphic principal 2-bundles. Pull-backs of principal 2-bundles can be defined via the pull-backs of the transition functions.

\subsection{Connections and curvatures on principal 2-bundles}

Recall that a connection $\nabla$ on a principal $\sG$-bundle $E$ over some manifold $M$ can be defined by a $\frg$-valued 1-form on the total space of $E$, where $\frg={\sf Lie}(\sG)$. This 1-form can be pulled back to the patches $U_a$, which yields $\frg$-valued 1-form potentials $A_a$. From the potentials, we derive curvature 2-forms $F_a=\dd A_a+A_a\wedge A_a$.

Gauge transformations are given by sections of the automorphism bundle of $E$ and are locally of the form
\begin{equation}
 A_a\ \mapsto\ \tilde{A}_a\ :=\ g_a^{-1} A_a g_a+ g^{-1}_a\dd g_a\eand F_a\ \mapsto\ \tilde{F}_a\ :=\ g_a^{-1} F_a g_a~,
\end{equation}
where $g_a:U_a\to \sG$.  On overlaps of patches, the potential 1-forms (and correspondingly the curvatures) are connected by patching relations induced by the transition functions, which take the form of gauge transformations restricted to the overlaps $U_a\cap U_b$:
\begin{equation}
 A_b\ =\ g_{ab}^{-1}A_ag_{ab}+g_{ab}^{-1}\dd g_{ab}\eand F_b\ =\ g_{ab}^{-1}F_ag_{ab}~,
\end{equation}
where, as before, the $g_{ab}$ are the transition functions.

Consider now a principal 2-bundle $E$ with a strict structure Lie 2-group $(H\overset{\sft}{\to}\sG,\acton)$ and the corresponding strict Lie 2-algebra $(\frh\overset{\sft}{\to}\frg,\acton)$. A connective structure on $E$ is given by a set of local potentials $(A_a,B_a)$, where $A_a$ are local 1-forms taking values in $\frg$, while $B_a$ are local 2-forms taking values in $\frh$. We define the corresponding curvatures according to
\begin{equation}\label{eq:DefOfFH}
 F_a\ :=\ \dd A_a+A_a\wedge A_a\eand H_a\ :=\ \nabla B_a\ :=\ \dd B_a+A_a\acton B_a~.
\end{equation}
Notice that here we abuse notation slightly by including the wedge product between differential forms into the action of `$\acton$'. Concretely, we have for any $\omega\in\Omega^r\otimes \frg$ and $\rho,\rho'\in \Omega^s\otimes \frh$
\begin{equation}\label{eq:CombConForms}
\begin{aligned}
  \sft(\omega\acton \rho)\ &=\ \omega\wedge\sft(\rho)-(-1)^{rs}\sft(\rho)\wedge\omega~,\\
  \sft(\rho)\acton \rho'\ &=\ \rho\wedge\rho'-(-1)^{rs}\rho'\wedge\rho~.
\end{aligned}
\end{equation}

Roughly speaking, the 2-form potentials are responsible for the parallel transport along a surface, while the 1-form potentials are responsible for the parallel transport along the boundary of the surface. It has been shown \cite{Girelli:2003ev,Baez:2004in,Baez:0511710} that for the parallel transport along surfaces to be reparametrisation invariant, the so-called {\em fake curvature} has to vanish:
\begin{equation}\label{eq:fakecurvature}
 \CF_a\ :=\ F_a-\sft(B_a)\ =\ 0~.
\end{equation}
This equation implies that $H_a$ obeys a Bianchi identity, that is,
\begin{subequations}
\begin{equation}\label{eq:Bianchi}
 \nabla H_a\ =\ F_a\acton B_a\ =\ \sft(B_a)\acton B_a\ =\ B_a\wedge B_a-B_a\wedge B_a\ =\ 0~,
 \end{equation}
where in the last step we have used \eqref{eq:CombConForms}. It also implies that
\begin{equation}
  \sft(H_a)\ =\ 0~.
\end{equation}
\end{subequations}
We shall refer to the equations \eqref{eq:DefOfFH} and \eqref{eq:fakecurvature} as non-Abelian tensor field equations.

Gauge transformations of a connective structure on a principal 2-bundle $E$ are given by sections of the automorphism 2-bundle of $E$. Locally, they are given by $\sG$-valued functions $g_a$ and $\frh$-valued 1-forms $\Lambda_a$. Their action on the potentials and curvatures reads as
\begin{equation}\label{eq:Space-time-GT}
\begin{aligned}
 A_a\ &\mapsto\ \tilde{A}_a\ :=\ g_a^{-1} A_a g_a+g_a^{-1} \dd g_a-\sft(\Lambda_a)~,\\
 B_a\ &\mapsto\ \tilde{B}_a\ :=\ g_a^{-1}\acton B_a -\tilde{A}_a\acton\Lambda_a-\dd \Lambda_a-\Lambda_a\wedge \Lambda_a~,\\
 F_a\ &\mapsto\ \tilde{F}_a\ \,:=\ g_a^{-1} F_a g_a-\sft(\dd\Lambda_a+\Lambda_a\wedge\Lambda_a)-\sft(\Lambda_a)\wedge \tilde{A}_a-\tilde{A}_a\wedge \sft(\Lambda_a)~,\\
 H_a\ &\mapsto\ \tilde{H}_a\ :=\ g_a^{-1}\acton H_a-(\tilde F_a-\sft(\tilde B_a))\acton \Lambda_a~.
\end{aligned}
\end{equation}
Note that the fake curvature \eqref{eq:fakecurvature} is covariant under these gauge transformations, that is, $\CF_a\mapsto g^{-1}_a\CF_a g_a$. Moreover, upon imposing $\CF_a=0$, we realise that the transformation of $H_a$ simplifies to $H_a\mapsto g_a^{-1}\acton H_a$.

The potential forms on different patches are again related by gauge transformations restricted to the overlaps. Here, the transformation is induced by the transition function $g_{ab}$ together with an $\frh$-valued 1-form $\Lambda_{ab}$ such that
\begin{equation}
\begin{aligned}
 A_b&\ =\ g^{-1}_{ab} A_a g_{ab}+g^{-1}_{ab} \dd g_{ab}-\sft(\Lambda_{ab})~,\\
 B_b&\ = \ g^{-1}_{ab}\acton B_a -A_b\acton\Lambda_{ab}-\dd \Lambda_{ab}-\Lambda_{ab}\wedge \Lambda_{ab}~
\end{aligned}
\end{equation}
on $U_a\cap U_b$.
The 1-forms $\Lambda_{ab}$ obey $\Lambda_{aa}=0$ and, on   $U_a\cap U_b\cap U_c$, satisfy the following cocycle condition:
\begin{equation}
 \Lambda_{ac}\ =\ \Lambda_{bc}+g_{bc}^{-1}\acton\Lambda_{ab}-g_{ac}^{-1}\acton(h_{abc}\nabla_ah_{abc}^{-1})~.
\end{equation}
This formula can be derived, for instance, by considering the chain of transformations $B_a\to B_b\to B_c\to B_a$ via the above patching conditions.

\section{Twistor description of self-dual tensor fields}\label{sec:PWbos}

So far we have considered principal 2-bundles equipped with connective structures on general manifolds.  In this section, we wish to present a twistor space interpretation of the non-Abelian {\it self-dual} tensor field equations on six-dimensional flat space-time $M^6$, that is,
\begin{equation}\label{eq:SDTFE}
 F\ =\ \dd A+A\wedge A\ =\ \sft(B)~,\quad H\ =\ \nabla B\ =\ \dd B+A\acton B~,\eand H\ =\ {\star H}~.
\end{equation}
By virtue of \eqref{eq:DefOfFH} and \eqref{eq:fakecurvature}, we then have that
\begin{equation}
 \nabla{\star H}\ =\ 0~.
 \end{equation}
Note that $H=\star H$ transforms covariantly under gauge transformations since $F=\sft(B)$.

The twistor space suitable to describe chiral theories is the space that parametrises totally null 3-planes in complexified space-time $M^6:=\FC^6$. It was introduced in \cite{Penrose:1985jw,Penrose:1986ca,Hughston:1986hb}  and used recently in  \cite{Saemann:2011nb,Mason:2011nw}\footnote{See \cite{Baston:1989,Chatterjee:1998} for an earlier account.} to describe, amongst other things, the Abelian self-dual tensor field equation. After a brief review of the construction of this twistor space, which we denote by $P^6$, we shall establish a Penrose--Ward transform  between equivalence classes of certain holomorphic principle 2-bundles over $P^6$ and gauge equivalence classes of solutions to \eqref{eq:SDTFE}. As common to many twistor constructions, we shall work in a complexified setting but reality conditions leading to Minkowski or Kleinian signature can be imposed at any stage of the construction.\footnote{See e.g.~Appendix A of \cite{Saemann:2011nb} for more details on this point.}

\subsection{Twistor space}

We have an identification of the tangent bundle $T_{M^6}\cong S\wedge S$ with the anti-symmetric tensor product of the (rank-4) bundle $S$ of anti-chiral spinors. Hence, we may coordinatise space-time $M^6$ by $x^{AB}=-x^{BA}$, where $A,B,\ldots=1,\ldots,4$ are spinor indices. Indices may be raised and lowered by the Levi-Civita symbol $\frac12\varepsilon_{ABCD}$, that is, $x_{AB}=\frac12\varepsilon_{ABCD}x^{CD}$. We also use partial derivatives with respect to $x^{AB}$ which we denote by $\partial_{AB}=\tfrac12\varepsilon_{ABCD}\partial^{CD}$ and $\partial_{AB}x^{CD}=\delta_{[A}^C\delta_{B]}^D$.\footnote{Brackets denote normalised anti-symmetrisation of the enclosed indices while we shall use parantheses to denote normalised symmetrisation.}

To define the twistor space $P^6$, we first consider the projectivisation $F^9:=\PP(S^\vee)\cong\FC^6\times\PP^3$ of the dual of $S$. We coordinatise $F^9$ by $(x^{AB},\lambda_A)$, where the $\lambda_A$ are homogenous coordinates on $\PP^3$, and refer to it as the {\em correspondence space}. On $F^9$ we consider the {\it twistor distribution} $D:=\mbox{span}\{V^A\}$ that is generated by the vector fields $V^A:=\lambda_B\partial^{AB}$; notice that  $\lambda_AV^A=0$ and hence, $D$ is rank-3 distribution. Since $D$ is integrable, we have a foliation of $F^9$ by three-dimensional complex manifolds. Finally, the twistor space $P^6:=F^9/D$ is  obtained as the quotient of the correspondence space by the twistor  distribution. Altogether, we arrive at a double fibration:
\begin{equation}\label{eq:DoubleFibration}
 \begin{picture}(50,40)
  \put(0.0,0.0){\makebox(0,0)[c]{$P^6$}}
  \put(64.0,0.0){\makebox(0,0)[c]{$M^6$}}
  \put(34.0,33.0){\makebox(0,0)[c]{$F^9$}}
  \put(7.0,18.0){\makebox(0,0)[c]{$\pi_1$}}
  \put(55.0,18.0){\makebox(0,0)[c]{$\pi_2$}}
  \put(25.0,25.0){\vector(-1,-1){18}}
  \put(37.0,25.0){\vector(1,-1){18}}
 \end{picture}
\end{equation}
Clearly, the projection $\pi_2$ is the trivial projection. To understand the projection $\pi_1$, we point out that $P^6$ can also be viewed as a hypersurface in $\PP^7\setminus\PP^3$. Concretely, if we let  $(z^A,\lambda_A)$ with $(\lambda_A)\neq(0,0,0,0)$ be homogeneous coordinates on  $\PP^7\setminus\PP^3$, then $P^6$ is given by the zero locus
\begin{equation}\label{eq:quadric}
 z^A\lambda_A\ =\ 0~.
\end{equation}
This makes clear that the projection $\pi_1$  in \eqref{eq:DoubleFibration} is given by
\begin{equation}
\pi_1:(x^{AB},\lambda_A)\ \mapsto\ (z^A,\lambda_A)\ =\ (x^{AB}\lambda_B,\lambda_A)~.
\end{equation}

Because of the {\it incidence relation},
 \begin{equation}\label{eq:incidence}
 z^A\ =\ x^{AB}\lambda_B~,
\end{equation}
any point $x\in M^6$ in space-time corresponds to a three-dimensional complex manifold $\hat x:=\pi_1(\pi_2^{-1}(x))\hookrightarrow P^6$ in twistor space which is bi-holomorphic to $\PP^3$. Conversely, for any point $p:=(z,\lambda)\in P^6$ in twistor space, we find a totally null 3-plane $\pi_2(\pi_1^{-1}(p))\hookrightarrow M^6$ in space-time given by
\begin{equation}\label{eq:IncidenceSolution}
 x^{AB}\ =\ x_0^{AB}+\varepsilon^{ABCD}\mu_C\lambda_D~,
\end{equation}
where $x_0^{AB}$ is a particular solution to \eqref{eq:incidence} and $\mu_A$ is defined modulo terms proportional to $\lambda_A$.

\subsection{Penrose--Ward transform}

Having recalled the basic ingredients, we shall now describe non-Abelian self-dual tensor fields via twistor geometry. To this end, let $\hat{\mathfrak{U}}=\{\hat U_a\}$ be a sufficiently fine open (Stein) cover of $P^6$ and let $\hat E\to P^6$ be a topologically trivial holomorphic principal 2-bundle with a strict structure Lie 2-group $(\sH\overset{\sft}{\to}\sG,\acton)$ over twistor space. As discussed in the previous section,  $\hat E$ can be described by a collection of holomorphic transition functions  $\hat g_{ab}:\hat U_a\cap \hat U_b\to \sG$ with $\hat g_{aa}=\unit$ and  $\hat h_{abc}:\hat U_a\cap \hat U_b\cap \hat U_c\to \sH$  with $\hat h_{aaa}=\hat h_{aab}=\hat h_{abb}=\unit$ subject to the cocycle condition \eqref{eq:cocylce} and modulo the equivalence relation \eqref{eq:equivalence}.  Note that because $\hat E$ is assumed to be topologically trivial, the transition functions $\hat g_{ab}$ and $\hat h_{abc}$ are smoothly (but {\it not} holomorphically) cohomologous to one: $\hat g_{ab}\sim\unit$ and $\hat h_{abc}\sim\unit$. In
spirit of the Ward construction \cite{Ward:1977ta} of self-dual Yang--Mills fields in four dimensions, we shall also assume that $\hat E$ becomes holomorphically trivial on any complex projective 3-space $\hat x=\pi_1(\pi_2^{-1}(x))\hookrightarrow P^6$ for $x\in M^6$. Put differently, upon restriction to $\hat x\cong\PP^3$, the transition functions  $\hat g_{ab}$ and $\hat h_{abc}$ are holomorphically (not just smoothly) cohomologous to one: $\hat g_{ab}|_{\hat x}\sim\unit$ and $\hat h_{abc}|_{\hat x}\sim\unit$.

Let us now pull back $\hat E$ via the projection $\pi_1$ to obtain a holomorphic principle 2-bundle $E\to F^9$ over the correspondence space, $E:=\pi_1^*\hat E$. If we let $\mathfrak{U}=\{U_a:=\pi_1^{-1}(\hat U_a)\}$ be the induced covering of $F^9$, then $E$ is described by the holomorphic transition functions $g_{ab}:=\pi_1^*\hat g_{ab}$ and $h_{abc}:=\pi_1^*\hat h_{abc}$. If we let $\dd_{\pi_1}$ be the relative exterior derivative along the fibration $\pi_1:F^9\to P^6$, that is, $\dd_{\pi_1}$ only contains the vector fields $V^A=\lambda_B\partial^{AB}$ generating the twistor distribution, then we have
\begin{equation}
 \dd_{\pi_1}g_{ab}\ =\ 0\quad\Leftrightarrow\quad V^Ag_{ab}\ =\ 0\eand  \dd_{\pi_1}h_{abc}\ =\ 0\quad\Leftrightarrow
 \quad V^Ah_{abc}\ =\ 0
\end{equation}
by the definition of a pull-back. Furthermore, because of our assumption of $\hat E$ being holomorphically trivial on any $\hat x$, the principle 2-bundle $E$ is holomorphically trivial on all of $F^9$ and hence, $g_{ab}$ and $h_{abc}$ are holomorphically cohomologous to one: $g_{ab}\sim\unit$ and $h_{abc}\sim\unit$. Explicitly, this means that $g_{ab}$ and $h_{abc}$ can be split as
\begin{subequations}\label{eq:Splitting}
\begin{eqnarray}
  g_{ab}&\!=\!&\sft(h_{ab}^{-1})g_ag_b^{-1}~,\label{eq:Splitting-g}\\
  h_{abc}&\!=\!&h_{ac}^{-1}h_{ab}(g_{ab}\acton h_{bc})~,\label{eq:Splitting-h}
\end{eqnarray}
\end{subequations}
as follows from \eqref{eqTriv2B}. Here, the maps $g_a:U_a\to \sG$ and $h_{ab}:U_a\cap U_b\to \sH$ are both holomorphic (and $h_{aa}=\unit$), however, $\dd_{\pi_1}g_a\neq 0$ and $\dd_{\pi_1}h_{ab}\neq 0$ in general.

This latter fact allows us to introduce relative differential 1-forms by setting
\begin{equation}
 a_a\ :=\ g_a^{-1}\dd_{\pi_1} g_a\eon U_a\eand b_{ab}\ :=\ g_a^{-1}\acton (\dd_{\pi_1}h_{ab}\,h_{ab}^{-1})\eon U_a\cap U_b~,
\end{equation}
i.e.~$a_a$ and $b_{ab}$ have components only along the fibration $\pi_1:F^9\to P^6$. In the following, we shall denote the sheaf of relative differential $r$-forms on $F^9$ by $\Omega_{\pi_1}^r$. Moreover,
\begin{equation}\label{eq:bab}
 \dd_{\pi_1}a_a+a_a\wedge a_a\ =\ 0\eand \dd_{\pi_1}b_{ab}-b_{ab}\wedge b_{ab}+a_a\acton b_{ab}\ =\ 0
\end{equation}
as a short calculation reveals. Furthermore, from \eqref{eq:Splitting-g} it follows  that  the $a_a$ are patched together by
\begin{equation}\label{eq:Patching-A}
   a_a\ =\ a_b+\sft(b_{ab})\eon U_a\cap U_b~,
\end{equation}
while from \eqref{eq:Splitting-h}, we find
\begin{equation}\label{eq:cohomH1}
 b_{ab}+b_{bc}+b_{ca}\ =\ 0\eon U_a\cap U_b\cap U_c~,
\end{equation}
and therefore, together with $b_{aa}=0$, we have $b_{ab}=-b_{ba}$. Equation \eqref{eq:cohomH1} implies that the $b_{ab}$ define an element of the cohomology group $H^1(F^9,\Omega_{\pi_1}^1\otimes\mathfrak{h})$. However, this group is zero\footnote{This simply follows from the fact that $H^1(F^9,\Omega_{\pi_1}^1)$ vanishes \cite{Saemann:2011nb}.} and hence, we can split $b_{ab}$ as
\begin{equation}\label{eq:Splitting-b}
 b_{ab}\ =\ b_a-b_b~.
\end{equation}
Furthermore, from $a_a-a_b=\sft(b_a)-\sft(b_b)$, we find that the relative 1-forms
\begin{equation}\label{eq:DefOfA}
 A_a\ :=\ a_a-\sft(b_a)
\end{equation}
are globally defined, that is, $A_a=A_b$ on $U_a\cap U_b$. Thus, there exists a globally defined relative 1-form $A_{\pi_1}$ with $A_a=A_{\pi_1}|_{U_a}$. Next we define a collection of relative 2-forms by
\begin{equation}\label{eq:DefOfB}
  B_a\ :=\ -(\dd_{\pi_1} b_a-b_a\wedge b_a+a_a\acton b_a)~.
  \end{equation}
 Using \eqref{eq:bab}, \eqref{eq:Patching-A}, and \eqref{eq:Splitting-b}, one can show that $B_a=B_b$ on $U_a\cap U_b$. Hence, the $B_a$ define a global relative 2-form $B_{\pi_1}$ on the correspondence space with $B_a=B_{\pi_1}|_{U_a}$.  The definitions \eqref{eq:DefOfA} and \eqref{eq:DefOfB} then yield
 \begin{equation}
 F_a \ :=\ \dd_{\pi_1} A_a+A_a\wedge A_a\ =\ \sft(B_a)\eand
 H_a \ :=\ \dd_{\pi_1}B_a+A_a\acton B_a\ =\ 0~.
\end{equation}
Clearly, $F_a=F_b$ and likewise $H_a=H_b$ since $A_a=A_b$ and $B_a=B_b$ on $U_a\cap U_b$ and hence, $F_a=F_{\pi_1}|_{U_a}$ and $H_a=H_{\pi_1}|_{U_a}$.

Note that the splitting  \eqref{eq:Splitting-g} is not unique as we can always perform $g_a\mapsto g_a g$ for some global holomorphic $\sG$-valued function $g$. Likewise the splitting \eqref{eq:Splitting-b} is unique only up to shifts by some $\mathfrak{h}$-valued global relative 1-form $\Lambda_{\pi_1}$, that is, $b_a\mapsto b_a+\Lambda_{\pi_1}$. Altogether,  the $a_a$ and $b_a$ are unique up to\footnote{Note that we  have the additional freedom $g_a\mapsto g_a\sft(h_a)$ and $h_{ab}\mapsto (g_a\acton h_ah_b^{-1})h_{ab}$ for some smooth $h_a:U_a\to\sH$ as these transformations (also known as modifications) leave the splitting \eqref{eq:Splitting} invariant. Such transformations will then be reflected in corresponding transformations of the 1-forms $a_a$ and $b_a$. However, the global forms $A_{\pi_1}$ and $B_{\pi_1}$ are invariant under such transformations, and, hence, these transformations will not lead to any additional space-time gauge transformations (see below).}
\begin{equation}
\begin{aligned}
  a_a\ \mapsto\ \tilde a_a\ :=\ g^{-1} a_a g+g^{-1} \dd_{\pi_1} g\eand
  b_a\ \mapsto\ \tilde b_a\ :=\ g^{-1}\acton b_a+\Lambda_{\pi_1}~.
  \end{aligned}
\end{equation}
Under these transformations, $A_a$ and $B_a$ and thus $A_{\pi_1}$ and $B_{\pi_1}$ behave as
\begin{equation}\label{eq:GT-AB}
\begin{aligned}
  A_{\pi_1}\ &\mapsto\ \tilde A_{\pi_1}\ =\ g^{-1} A_{\pi_1} g+g^{-1} \dd_{\pi_1} g -\sft(\Lambda_{\pi_1})~,\\
  B_{\pi_1}\ &\mapsto\ \tilde B_{\pi_1}\ =\ g^{-1}\acton B_a-\tilde A_{\pi_1}\acton \Lambda_{\pi_1}-\dd_{\pi_1}\Lambda_{\pi_1}-\Lambda_{\pi_1}\wedge\Lambda_{\pi_1}~.
   \end{aligned}
\end{equation}
As it should be, the relation $F_{\pi_1}=\sft(B_{\pi_1})$ behaves covariantly under \eqref{eq:GT-AB}, that is, $\tilde F_{\pi_1}=\sft(\tilde B_{\pi_1})$. We shall see momentarily that these transformations will correspond to space-time gauge transformations.

In summary, we have constructed a {\it relative connective structure} on the holomorphically trivial principal 2-bundle $E\to F^9$ consisting of a relative 1-form $A_{\pi_1}$ and relative 2-form $B_{\pi_1}$ (both defined globally) such that $F_{\pi_1}=\dd_{\pi_1}A_{\pi_1}+A_{\pi_1}\wedge A_{\pi_1}=\sft(B_{\pi_1})$ and $H_{\pi_1}=\dd_{\pi_1} B_{\pi_1}+A_{\pi_1}\acton B_{\pi_1}=0$. Because of the last equation, the relative connective structure is said to be {\it 2-flat}. The final step in our construction is to push this relative connective structure on $E$ down to space-time to obtain a holomorphic principal 2-bundle $E'\to M^6$. Clearly, $E'$ will be holomorphically trivial but as we shall see momentarily, the connective structure $(A,B)$ on $E'$ will be self-dual.

Using the isomorphisms between global relative differential $r$-forms on the correspondence space and spinor fields on space-time as obtained in   \cite{Saemann:2011nb} (see also \cite{Mason:2011nw}), we may expand $(A_{\pi_1}, B_{\pi_1})$ and $(F_{\pi_1},H_{\pi_1})$ as
\begin{equation}\label{eq:RelFormExp}
\begin{aligned}
 A_{\pi_1}\ &=\ e_{[A}\lambda_{B]}\, A^{AB}~,\\
 B_{\pi_1}\ &=\ e_A\wedge e_B\lambda_C\, \varepsilon^{ABCD} B_D{}^{E}\lambda_{E}~,\\
 F_{\pi_1}\ &=\  -\tfrac14 e_A\wedge e_B\lambda_C\, \varepsilon^{ABCD} F_D{}^{E}\lambda_{E}~,\\
 H_{\pi_1}\ &=\  \ -\tfrac13 e_A\wedge e_B\wedge e_C\lambda_D\varepsilon^{ABCD}\,H^{EF}\lambda_E\lambda_F~,
 \end{aligned}
\end{equation}
where all the $\lambda$-dependence has been made explicit. Here, we used the relative 1-forms $e_A$ of homogeneity $-1$, which combine with $V^A$ to the relative exterior derivative $\dd_{\pi_1}=e_A V^A$. The $e_A$ are defined modulo terms proportional to $\lambda_A$ since $\lambda_A V^A=0$. The above expansions of $(A_{\pi_1},B_{\pi_1})$ and $(F_{\pi_1},H_{\pi_1})$ reflect this property.  Furthermore, $A_{AB}=-A_{BA}=\frac12\varepsilon_{ABCD}A^{CD}$, $H_{AB}=H_{BA}$ and both $B_A{}^B$ and $F_A{}^B$ are trace-less. The pre-factors were inserted for later convenience.  Then from $F_{\pi_1}=\dd_{\pi_1}A_{\pi_1}+A_{\pi_1}\wedge A_{\pi_1}=\sft(B_{\pi_1})$ and $H_{\pi_1}=\dd_{\pi_1} B_{\pi_1}+A_{\pi_1}\acton B_{\pi_1}=0$, we find
\begin{subequations}\label{eq:EoM-ST}
\begin{equation}
 F_A{}^B\ =\ \sft(B_A{}^B)\eand H^{AB}\ =\ \nabla^{C(A}B_C{}^{B)}\ =\ 0~,
\end{equation}
with

\begin{equation}
\begin{aligned}
 F_A{}^B\ &:=\ \partial^{BC} A_{CA}-\partial_{CA}A^{BC}+[A^{BC},A_{CA}]~,\\
  \nabla^{C(A}B_C{}^{B)}\ &:=\ \partial^{C(A}B_C{}^{B)}+A^{C(A}\acton B_C{}^{B)}~.
  \end{aligned}
\end{equation}
\end{subequations}
In general, a 3-form $H=\nabla B=\dd B+A\acton B$ on space-time reads in spinor notation as
\begin{equation}
  (H_{AB},H^{AB})\ =\ (\nabla_{C(A}B_{B)}{}^{C}, \nabla^{C(A}B_C{}^{B)})~,
\end{equation}
where $H_{AB}$ and $H^{AB}$ represent the self-dual and anti-self-dual parts of $H$, respectively.
Therefore, using the 1-form $A_{AB}$ and the 2-form $B_A{}^B$ obtained in \eqref{eq:EoM-ST}, we find a 3-form $H$
\begin{equation}
  (H_{AB},H^{AB})\ =\ (\nabla_{C(A}B_{B)}{}^{C}, \nabla^{C(A}B_C{}^{B)})\ =\ (\nabla_{C(A}B_{B)}{}^{C},0)~,
\end{equation}
which is self-dual, i.e.~$H={\star H}$. Hence, $\nabla{\star H}=\nabla H=0$. In spinor notation, this is
\begin{equation}
 \nabla^{AC}H_{CB}\ =\ 0~.
\end{equation}
We point out that the equations of motion  \eqref{eq:EoM-ST} constitute, in fact, a non-linear set of differential equations for $B$ since roughly speaking, $A$ is determined by $B$ via the fake curvature equation. In that sense, $A$ does not contain additional (physical) degrees of freedom and we end up with a self-dual 3-form $H$ determined by the 2-form potential $B$. In addition, note that since the gauge parameter $g$ appearing in \eqref{eq:GT-AB} is holomorphic and globally defined it can only depend on $x^{AB}$ (since $\PP^3$ is compact). Likewise, the gauge parameter $\Lambda_{\pi_1}$ is holomorphic and defined globally, so there is a similar expansion for $\Lambda_{\pi_1}$ as for $A_{\pi_1}$. Thus, the gauge transformations  \eqref{eq:GT-AB}  reduce to the space-time gauge transformations displayed in   \eqref{eq:Space-time-GT}. That is, the non-uniqueness in the splittings of the transition functions results in gauge freedom of the space-time fields.

Conversely, any  such holomorphic principal 2-bundle over the twistor space arises from solutions to \eqref{eq:EoM-ST} and the expansions \eqref{eq:RelFormExp}. Altogether, we may summarise our above discussion in the following theorem:

\pagebreak[4]
{\theorem
There is a bijection between
\begin{itemize}
\item[\rm {(i)}] equivalence classes of topologically trivial holomorphic principal 2-bundles over the twistor space $P^6$ that are holomorphically trivial when restricted to any complex projective 3-space $\hat x=\pi_1(\pi_2^{-1}(x))\hookrightarrow P^6$ for $x\in M^6$ and
\item[{\rm (ii)}] gauge equivalence classes of (complex holomorphic) solutions to the non-Abelian self-dual tensor field equations \eqref{eq:SDTFE} on space-time $M^6$.
\end{itemize}
}

\section{Supersymmetric extension}\label{sec:PWSUSY}

In this section, we would like to extend the above discussion supersymmetrically. In particular, we shall first introduce $\CN=(n,0)$ superspace for $n=1,2$, discuss its associated supertwistor space, and eventually derive a set of  non-Abelian constraint equations involving the tensor  multiplet.

\subsection{Supertwistor space}

To discuss $\CN=(n,0)$ supersymmetry, let us extend space-time $M^6$ by $8n$ fermionic directions and consider $\CN=(n,0)$ superspace $M^{6|8n}\cong \FC^{6|8n}:=\FC^6\oplus\Pi\FC^{8n}$ with coordinates $(x^{AB},\eta^A_I)$, where $x^{AB}$ are the usual Gra{\ss}mann even (bosonic) coordinates and $\eta^A_I$ are the Gra{\ss}mann odd (fermionic) coordinates with R-symmetry indices $I,J,\ldots=1,\ldots,2n$. Here, $\Pi$ is the Gra{\ss}mann parity changing operator. We then introduce the derivatives
\begin{equation}
  P_{AB}\ :=\ \der{x^{AB}}\eand D^I_A\ :=\ \der{\eta^A_I}-2\Omega^{IJ}\eta_J^B\der{x^{AB}}~,
\end{equation}
which satisfy the anti-commutation relations
\begin{equation}
 \{D^I_A,D^J_B\}\ =\ -4\Omega^{IJ}P_{AB}~,
\end{equation}
where $\Omega:=(\Omega^{IJ})$ is an $\sSp(n)$-invariant $2n\times 2n$ matrix. Here, $\sSp(1)\cong \sSU(2)$ and $\sSp(2)\cong\mathsf{USp}(4)\subset \sSp(4,\FC)$ are the R-symmetry groups of the superconformal groups $\sOSp(2,6|2n)$ in six dimensions. The group $\sSp(2)$ is defined as the elements of $\sSU(4)$ leaving $\Omega$ invariant. It is therefore given by the intersection $\sSU(4)\cap \sSp(4,\FC)$. We may choose $\Omega$ to read as\footnote{Note that we are working in a complexified setting, and one could therefore wonder about the existence of $\Omega$ for the complexification of $\sSp(n)$. As we can always impose reality conditions to restrict to the real case, we ignore this point in the following.}
\begin{equation}
 \Omega\ =\ \diag(\underbrace{\varepsilon,\ldots,\varepsilon}_{n\rm{-times}})\ewith\varepsilon\ :=\ \left(\begin{array}{cc} 0 & 1 \\ -1& 0 \end{array}\right).
\end{equation}

The correspondence space is defined in a similar manner as before, that is, $F^{9|8n}\cong\FC^{6|8n}\times\PP^3$ and equipped with the coordinates $(x^{AB},\eta^A_I,\lambda_A)$. Then the  twistor distribution generalises to
\begin{equation}
  D\ :=\ \mbox{span}\{ V^A, V^{IAB}\}\ewith
  V^A\ :=\ \lambda_B\partial^{AB}\eand
  V^{IAB}\ :=\ \tfrac12\varepsilon^{ABCD}\lambda_C D^I_D~,
\end{equation}
which is a rank-$3|6n$ distribution since $\lambda_A V^A=0=\lambda_B V^{IAB}$. One can check that $D$ is integrable. Therefore, we may define supertwistor space by the quotient $P^{6|2n}:=F^{9|8n}/D$ so that
\begin{equation}\label{eq:superDoubleFibration}
 \begin{picture}(50,40)
  \put(0.0,0.0){\makebox(0,0)[c]{$P^{6|2n}$}}
  \put(64.0,0.0){\makebox(0,0)[c]{$M^{6|8n}$}}
  \put(34.0,33.0){\makebox(0,0)[c]{$F^{9|8n}$}}
  \put(7.0,18.0){\makebox(0,0)[c]{$\pi_1$}}
  \put(55.0,18.0){\makebox(0,0)[c]{$\pi_2$}}
  \put(25.0,25.0){\vector(-1,-1){18}}
  \put(37.0,25.0){\vector(1,-1){18}}
 \end{picture}
\end{equation}
As in the purely bosonic case, $\pi_2$ is the trivial projection while $\pi_1$ acts as
\begin{equation}
\pi_1:(x^{AB},\eta^A_I,\lambda_A)\ \mapsto\ (z^A,\eta_I,\lambda_A)\ =\ ((x^{AB}+\Omega^{IJ}\eta^A_I\eta^B_J)\lambda_B,\eta_I^A\lambda_A,\lambda_A)~,
\end{equation}
so that the incidence relation reads as
\begin{equation}\label{eq:superincidence}
 z^A\ =\ (x^{AB}+\Omega^{IJ}\eta^A_I\eta^B_J)\lambda_B\eand
 \eta_I\ =\ \eta_I^A\lambda_A~.
\end{equation}
Furthermore, the quadric equation equation \eqref{eq:quadric} becomes
\begin{equation}\label{eq:superquadric}
 z^A\lambda_A-\Omega^{IJ}\eta_I\eta_J\ =\ 0~.
\end{equation}

As before, the incidence relation \eqref{eq:superincidence} establishes a relation between points and certain submanifolds. In particular, any point $x\in M^{6|8n}$ corresponds to a complex projective 3-space  $\hat x=\pi_1(\pi_2^{-1}(x))\hookrightarrow P^{6|2n}$ while for any point $p\in P^{6|2n}$ in twistor space we find a totally null $3|6n$-superplane $\pi_2(\pi_1^{-1}(p))\hookrightarrow M^{6|8n}$ given by
\begin{equation}\label{eq:IncidenceSolutionSuper}
\begin{aligned}
 x^{AB}\ &=\ x_0^{AB}+\varepsilon^{ABCD}\mu_C\lambda_D+2\Omega^{IJ}\varepsilon^{CDE[A}\lambda_C\theta_{IDE}\eta_0{}^{B]}_J~,\\
 \eta^A_I\ &=\ \eta_0{}^A_I+\varepsilon^{ABCD}\lambda_B\theta_{ICD}~.
 \end{aligned}
\end{equation}
Here, $(x_0^{AB},\eta_0{}^A_I)$ is a particular solution to \eqref{eq:superincidence} while the parameters $\mu_A$ and $\theta_{IAB}$ are defined modulo $\lambda_A$ which implies that we have $3|6n$ parameters in total parametrising a totally null $3|6n$-superplane.

\subsection{Penrose--Ward transform and constraint equations}

Let us now consider a topologically trivial holomorphic principal 2-bundle $\hat E\to P^{6|2n}$ with a strict structure Lie 2-group $(\sH\overset{\sft}{\to}\sG,\acton)$ over supertwistor space $P^{6|2n}$.\footnote{We could also work with supergroups at this stage.}  We shall assume that $\hat E$ becomes holomorphically trivial upon restriction to any projective 3-space $\hat x=\pi_1(\pi_2^{-1}(x))\hookrightarrow P^{6|2n}$. We may now follow the same steps as in the previous section to arrive at a holomorphically trivial principal 2-bundle $E\to F^{9|8n}$ on the correspondence space equipped with a 2-flat relative connective structure. As the derivation is essentially the same, we do not need  to repeat it here. The equations we find are
\begin{equation}\label{eq:superflatness}
\begin{aligned}
 F_{\pi_1}\ &=\ \dd_{\pi_1}A_{\pi_1}+A_{\pi_1}\wedge A_{\pi_1}\ =\ \sft(B_{\pi_1})~,\\
 H_{\pi_1}\ &=\ \dd_{\pi_1} B_{\pi_1}+A_{\pi_1}\acton B_{\pi_1}\ =\ 0~.
 \end{aligned}
\end{equation}

Now we would like to push these equations down to $M^{6|8n}$. Of course, the expansions of the relative differential forms $A_{\pi_1}$, $B_{\pi_1}$, $F_{\pi_1}$, and $H_{\pi_1}$ are more complicated due to the fermionic directions. In particular, the relative exterior derivative reads as
\begin{equation}
  \dd_{\pi_1}\ =\ e_A V^A+e_{IAB} V^{IAB}\ =\ e_{[A}\lambda_{B]}\partial^{AB}+e^{AB}_I\lambda_A D^I_B~,
\end{equation}
where the $e_A$ and $e_{IAB}=\frac12\varepsilon_{ABCD}e^{CD}_I$ are defined modulo terms proportional to $\lambda_A$. Then we have
\begin{equation}\label{eq:expansions}
\begin{aligned}
 A_{\pi_1}\ &=\ e_{[A}\lambda_{B]}\,  A^{AB}+e^{AB}_I\lambda_A\, A^I_B~,\\
 B_{\pi_1}\ &=\ -\tfrac14e_A\wedge e_B\lambda_C\, \varepsilon^{ABCD} B_D{}^{E}\lambda_{E}+\tfrac12 e_A\lambda_B\wedge e^{EF}_I\lambda_E\,\varepsilon^{ABCD}\,B_{CD}{}^I_F~+\\
   &\kern1cm+\tfrac12 e^{CA}_I\lambda_C\wedge e^{DB}_J\lambda_D\, B^{IJ}_{AB}~,\\
 F_{\pi_1}\ &=\ -\tfrac14 e_A\wedge e_B\lambda_C\, \varepsilon^{ABCD} F_D{}^{E}\lambda_{E}+\tfrac12 e_A\lambda_B\wedge e^{EF}_I\lambda_E\,\varepsilon^{ABCD}\,F_{CD}{}^I_F~+\\
   &\kern1cm+\tfrac12 e^{CA}_I\lambda_C\wedge e^{DB}_J\lambda_D\, F^{IJ}_{AB}~,\\
  H_{\pi_1}\ &=\  -\tfrac13 e_A\wedge e_B\wedge e_C\lambda_D\varepsilon^{ABCD}\,H^{EF}\lambda_E\lambda_F~+\\
 &\kern1cm -\tfrac14 e_A\wedge e_B\lambda_C\, \varepsilon^{ABCD}\wedge e^{EF}_I\lambda_E\, (H_{D}{}^G{}^I_F)_0\lambda_G~+\\
 &\kern1cm + \tfrac14 e_A\lambda_B\wedge e^{EF}_I\lambda_E\wedge e^{GH}_J\lambda_G\,\varepsilon^{ABCD}\,(H_{CD}{}^{IJ}_{FH})_0~+\\
 &\kern1cm +\tfrac16 e^{DA}_I\lambda_D\wedge e^{EB}_J\lambda_E\wedge e^{FC}_K\lambda_F\, H^{IJK}_{ABC}~,
 \end{aligned}
\end{equation}
where all the $\lambda$-dependence has been made explicit. Here, $(H_{A}{}^B{}^I_C)_0$ denotes the totally trace-less part of $H_{A}{}^B{}^I_C$ while $(H_{AB}{}^{IJ}_{CD})_0$ indicates the part of $H_{AB}{}^{IJ}_{CD}$ that does not contain $\varepsilon_{ABCD}$. In general, a differential 2-form $F$ on $M^{6|8n}$ has components
\begin{subequations}
\begin{equation}
  (F_A{}^B,F_{AB}{}^I_C, F^{IJ}_{AB})~,
 \end{equation}
 where $F_A{}^B$ is trace-less, while a differential 3-form $H$ on $M^{6|8n}$ has a priori the components
\begin{equation}
  \big(H_{AB},H^{AB},H_{A}{}^B{}^I_C,H_{AB}{}^{IJ}_{CD},H^{IJK}_{ABC}\big)~,
\end{equation}
\end{subequations}
where $H_{A}{}^B{}^I_C$ is trace-less over the $AB$ indices. Thus, \eqref{eq:superflatness} together with \eqref{eq:expansions} yield the following set of constraint equations on $M^{6|8n}$:
\begin{subequations}\label{eq:constrainteq}
\begin{equation}
  F_A{}^B\ =\ \sft(B_A{}^B)~,\quad F_{AB}{}^I_C\ =\ \sft(B_{AB}{}^I_C)~,\eand F^{IJ}_{AB}\ =\ \sft(B^{IJ}_{AB})~,
\end{equation}
and
\begin{equation}
\begin{aligned}
  H^{AB}\ &=\ 0~,\\
  H_{A}{}^B{}^I_C\ &=\ \delta^B_C\psi^I_A-\tfrac14\delta^B_A\psi^I_C~,\\
  H_{AB}{}^{IJ}_{CD}\ &=\ \varepsilon_{ABCD}\phi^{IJ}~,\\
  H^{IJK}_{ABC}\ &=\ 0~,
  \end{aligned}
\end{equation}
\end{subequations}
where $\psi^I_A$ is a fermionic spinor field and $\phi^{IJ}$ represents bosonic scalar fields. These equations arise since $H_{\pi_1}=0$ implies $(H_{A}{}^B{}^I_C)_0=0$  and $(H_{AB}{}^{IJ}_{CD})_0=0$. Explicitly, the curvatures $(F,H)$ are given in terms of the gauge potentials $(A,B)$ by
\begin{subequations}
\begin{equation}
\begin{aligned}
  F_A{}^B\ &=\ \partial^{BC} A_{CA}-\partial_{CA}A^{BC}+[A^{BC},A_{CA}]~,\\
  F_{AB}{}^I_C\ &=\ ~\partial_{AB}A^I_C-D^I_CA_{AB}+[A_{AB},A^I_C]~,\\
  F^{IJ}_{AB}\ &=\ D^I_AA^J_B+D^J_B A^I_A+\{A^I_A,A^J_B\}+4\Omega^{IJ}A_{AB}
  \end{aligned}
\end{equation}
and
\begin{equation}
\begin{aligned}
  H_{AB}\ &=\ \nabla_{C(A}B_{B)}^C\eand H^{AB}\ =\ \nabla^{C(A}B_C{}^{B)}~,\\
  H_{A}{}^B{}^I_C\ &=\ \nabla^I_CB_A{}^B-\nabla^{DB}B_{DA}{}^I_C+\nabla_{DA}B^{DB}{}^I_C~,\\
  H_{AB}{}^{IJ}_{CD}\ &=\ \nabla_{AB}B^{IJ}_{CD}-\nabla^I_C B_{AB}{}^J_D-\nabla^J_D B_{AB}{}^I_C-2\Omega^{IJ}(\varepsilon_{ABF[C} B_{D]}{}^F-\varepsilon_{CDF[A} B_{B]}{}^F)~,\\
  H^{IJK}_{ABC}\ &=\ \nabla^I_A B^{JK}_{BC}+\nabla^J_BB^{IK}_{AC}+\nabla^K_C B^{IJ}_{AB}\\
     &\kern1cm+4\Omega^{IJ}B_{AB}{}^K_C+4\Omega^{IK}B_{AC}{}^J_B+4\Omega^{JK}B_{BC}{}^I_A~.
  \end{aligned}
\end{equation}
\end{subequations}

Altogether, we arrive at the following theorem:

{\theorem
There is a bijection between
\begin{itemize}
\item[\rm {(i)}] equivalence classes of topologically trivial holomorphic principal 2-bundles over the supertwistor space $P^{6|2n}$ that are holomorphically trivial when restricted to any complex projective 3-space $\hat x=\pi_1(\pi_2^{-1}(x))\hookrightarrow P^{6|2n}$ for $x\in M^{6|8n}$ and
\item[{\rm (ii)}] gauge equivalence classes of (complex holomorphic) solutions to the constraint equations \eqref{eq:constrainteq} on the superspace $M^{6|8n}$.
\end{itemize}
}

Let us conclude with some remarks concerning the physical field content of our constraint equations. Under gauge transformations, the fields $(H_{AB},\psi^I_A,\phi^{IJ})$ transform on-shell as  $(H_{AB},\psi^I_A,\phi^{IJ})\mapsto g^{-1}\acton  (H_{AB},\psi^I_A,\phi^{IJ})$. Furthermore, at the linearised level, it is easy to see that they satisfy the (superspace) free field equations
\begin{equation}\label{eq:lineq}
 \partial^{AC}H_{CB}\ =\ 0~,~~~\dpar^{AB}\psi_A^I\ =\ 0~,\eand \dpar^{AB}\dpar_{AB}\phi^{IJ}\ =\ 0~.
\end{equation}
For $n=1$, the fields $(H_{AB},\psi^I_A,\phi^{IJ})$ constitute the $\CN=(1,0)$ tensor multiplet consisting of a self-dual 3-form, two spinors, and a scalar. The $\CN=(2,0)$ tensor multiplet consists of a self-dual 3-form, four spinors, and five scalars. It seems, however, that  for $n=2$ or, equivalently,  for $\CN=(2,0)$ supersymmetry, the twistor space constructions yields six scalars instead. This apparent problem can be resolved by inspecting the Bianchi identity $\nabla H=0$ for the self-dual 3-form. If and only if $n=2$, its purely fermionic part, $(\nabla H)_{ABCD}^{IJKL}=0$, together with the constraint equations \eqref{eq:constrainteq} yields the algebraic condition $\Omega_{IJ}\phi^{IJ}=0$, which, in turn, reduces the number of scalars by one. Thus, also for $\CN=(2,0)$ supersymmetry, we find the correct number of scalar fields.  Note that this is quite different from the ambitwistor construction of maximally supersymmetric Yang--Mills theory in four dimensions in the $\CN=4$ formalism.\footnote{Notice that the ambitwistor space sits naturally in $P^6$ \cite{Saemann:2011nb}.} In that approach, the twistor description yields too many scalar fields, as well, and an additional algebraic condition, which does {\it not} follow from the ambitwistor construction, has to be imposed by hand to restrict the field content to that of maximally supersymmetric Yang--Mills theory \cite{Witten:1978xx}.\footnote{This problem can be avoided by working in the $\CN=3$ formalism at the cost of not making manifest the full $\CN=4$ supersymmetry. Note that such a constraint is absent in the twistor description of $\CN=(1,1)$ super Yang--Mills theory in six dimensions \cite{Saemann:2012rr} as also in that case, the field content is correctly reproduced.}   

Summarising, we have constructed non-Abelian constraint equations containing the $\CN=(n,0)$ tensor multiplet, for $n=0,1,2$, which, at the linearised level, obeys \eqref{eq:lineq}.

\section{Supersymmetric non-Abelian self-dual strings}\label{sec:sds}

The (Abelian) self-dual string equation \cite{Howe:1997ue} follows directly from a dimensional reduction of the self-duality equation $H_{\rm 6d}=\star H_{\rm 6d}$ of a 3-form field strength $H_{\rm 6d}=\dd B_{\rm 6d}$ from six dimensions to four dimensions. Explicitly, we have $H_{\rm 4d}=\star \dd \phi$, where $H_{\rm 4d}=\dd B_{\rm 4d}$ and $\phi$ is a Higgs field that contains the component of $B_{6d}$ along the reduced directions. A further reduction to three dimensions yields the Bogomolny monopole equation. Self-dual strings can therefore be considered as M-theory lifts of magnetic monopoles.

Because dimensional reductions on space-time have usually rather straightforward interpretations on twistor space, we can easily derive the twistor description of non-Abelian supersymmetric self-dual strings from our above discussion. In the Abelian and bosonic case, the appropriate twistor space is the hyperplane twistor space constructed in \cite{Saemann:2011nb}. Here, it merely remains to supersymmetrically extend this twistor space and adapt the domain of the Penrose--Ward transform to holomorphic principal 2-bundles. We will be therefore rather concise in our description and refer to \cite{Saemann:2011nb} for most details.

On four-dimensional superspace $M^{4|8n}\cong \FC^{4|8n}$, we have coordinates $x^{\alpha\ald},\eta_I^\ald,\theta^{I \alpha}$, where $\alpha,\ald=1,2$ are spinor indices of $\sSL(2,\FC)$ and $I=1,\ldots,2n$ is the R-symmetry index. These coordinates arise from the ones on $M^{6|8n}$ in a straightforward manner; the R-symmetry index is raised using the $\sSp(n)$-invariant matrix $\Omega^{IJ}$.

The correspondence space $F^{6|8n}$ is defined as the product $M^{4|8n}\times \PP^1\times \PP^1$, which we coordinatise by $x^{\alpha\dot\beta},\eta_I^\ald,\theta^{I \alpha},\lambda_\ald$, and $\mu_\alpha$. To obtain twistor space, we divide by the twistor distribution, which is spanned by the vector fields
\begin{subequations}
\begin{equation}
 V_\alpha\ :=\ \lambda^{\dot\beta}\der{x^{\alpha\dot\beta}}~,~~~V_\ald\ :=\ \mu^\beta\der{x^{\beta\ald}}~,~~~V^I\ :=\ \lambda^\ald D_{\ald}^I~,~~~V_I\ :=\ \mu^\alpha D_{I\alpha}~,
\end{equation}
where
\begin{equation}
 D^I_{\ald}\ :=\ \der{\eta_I^\ald}-2\theta^{I\beta}\der{x^{\beta\ald}}\eand 
 D_{I\alpha}\ :=\ \der{\theta^{I\alpha}}+2\eta_I^{\dot\beta}\der{x^{\alpha\dot\beta}}~.
\end{equation}
\end{subequations}
As usual, we raise and lower the spinor indices $\alpha,\ald$ with the invariant $\eps$-tensors of $\sSL(2,\FC)$, e.g.\ $\lambda^\ald=\eps^{\ald\bed}\lambda_\bed$. Notice that $\lambda^\ald V_\ald=\mu^\alpha V_\alpha$. Altogether, we arrive at the following twistor correspondence, which is the supersymmetric extension of the hyperplane twistor space:
\begin{equation}\label{eq:superDoubleFibrationSDS}
 \begin{picture}(50,40)
  \put(0.0,0.0){\makebox(0,0)[c]{$P^{3|4n}$}}
  \put(64.0,0.0){\makebox(0,0)[c]{$M^{4|8n}$}}
  \put(34.0,33.0){\makebox(0,0)[c]{$F^{6|8n}$}}
  \put(7.0,18.0){\makebox(0,0)[c]{$\pi_1$}}
  \put(55.0,18.0){\makebox(0,0)[c]{$\pi_2$}}
  \put(25.0,25.0){\vector(-1,-1){18}}
  \put(37.0,25.0){\vector(1,-1){18}}
 \end{picture}
\end{equation}
Here, $\pi_2$ is the trivial projection while $\pi_1$ acts as
\begin{equation}
\pi_1:(x^{\alpha\dot\beta},\eta^\ald_I,\theta^{I\alpha},\lambda_\ald,\mu_\alpha)\ \mapsto\ (z,\eta_I,\theta^I,\lambda_\ald,\mu_\alpha)\ =\ (x^{\alpha\dot\beta}\mu_\alpha\lambda_{\dot\beta},{\eta}_I^\ald\lambda_\ald,{\theta}^{I\alpha}\mu_\alpha,\lambda_\ald,\mu_\alpha)~
\end{equation}
with the incidence relation
\begin{equation}
 z\ =\ x^{\alpha\dot\beta}\mu_\alpha\lambda_{\dot\beta}~,~~~
 \eta_I\ =\ {\eta}_I^\ald\lambda_\ald~,\eand
 \theta^I\ =\ {\theta}^{I\alpha}\mu_\alpha~.
\end{equation}
We can identify the twistor space $P^{3|4n}$ with the total space of the vector bundle $\CO(1,1)\oplus \big(\FC^{2n}\otimes(\Pi\CO(1,0)\oplus\Pi\CO(0,1))\big)$ over $\PP^1\times \PP^1$. One readily verifies that here, the geometric twistor correspondence is between points in $M^{4|8n}$ and sections of $P^{3|4n}$ as well as between points on $P^{3|4n}$ and supersymmetric, $3|4n$-dimensional hyperplanes in $M^{4|8n}$. This completes the description of the hyperplane supertwistor space.

The Penrose--Ward transform is now straightforwardly adapted from the twistor space $P^{6|2n}$ to the hyperplane supertwistor space $P^{3|4n}$. We start from a topologically trivial holomorphic principal 2-bundle $\hat{E}$ with a strict structure Lie 2-group $(\sH\overset{\sft}{\to}\sG,\acton)$ over $P^{3|4n}$ and pull it back along $\pi_1$. The \v Cech cocycles describing the pullback 2-bundle $\pi^*_1\hat{E}$ give rise to a relative connective structure along $\pi_1$. Expanding the corresponding differential forms in the homogeneous coordinates $\lambda_\ald$ and $\mu_\alpha$, we find the relevant field equations on $M^{4|8n}$. These are mere dimensional reductions of equations \eqref{eq:constrainteq}. Explicitly, we arrive at the connective structure $(A,B)$ of a trivial principal 2-bundle with a strict structure Lie 2-group $(\sH\overset{\sft}{\to}\sG,\acton)$ over $M^{4|8n}$ together with a Higgs field $\phi$ taking values in the Lie algebra of $\sH$, cf.\ \cite{Saemann:2011nb}. For the sake of brevity, we focus on the equations for the purely bosonic components of the connective structure; the other equations are easily derived from \eqref{eq:constrainteq}. In particular, we have the fake curvature conditions
\begin{subequations}
\begin{equation}
 f_{\alpha\beta}\ =\ \sft(B_{\alpha\beta})~,~~~f_{\ald\bed}\ =\ \sft(B_{\ald\bed})~,\eand \sft(\phi)\ =\ 0
\end{equation}
and the non-Abelian self-dual string equation
\begin{equation}
 \eps^{\bed\gad}\nabla_{\alpha\bed}B_{\ald\gad}-\eps^{\beta\gamma}\nabla_{\beta\ald}B_{\alpha\gamma}\ =\ \nabla_{\alpha\ald}\phi~.
\end{equation}
\end{subequations}

\section{Conclusions}\label{sec:conc}

We started from the twistor space $P^6$ that underlies the description of self-dual 3-forms in six dimensions in terms of holomorphic Abelian gerbes. We extended this twistorial description by considering holomorphic principal 2-bundles over $P^6$ subject to certain triviality conditions. This led to a bijection between equivalence classes of such 2-bundles and gauge equivalence classes of solutions to higher gauge theory for a self-dual 3-form curvature.  We then considered a supersymmetric extension $P^{6|2n}$ of the twistor space $P^6$. Here, we derived a set of $\CN=(n,0)$ supersymmetric non-Abelian constraint equations \eqref{eq:constrainteq} on superspace that contain the tensor multiplet. By virtue of the Penrose--Ward transform, all solutions to these constraint equations were shown to be given by holomorphic principal 2-bundles over $P^{6|2n}$. In that sense, \eqref{eq:constrainteq} constitutes a (classically) integrable theory of interacting 2-forms including scalars and fermions. From our twistor description of a higher gauge theory containing the tensor multiplet, we could directly derive a corresponding picture for non-Abelian supersymmetric self-dual strings.

The most obvious open problem to address is the in-depth analysis of the constraint equations we found on superspace and the extraction of the corresponding equations of motion on space-time. That is, we have to develop the six-dimensional analogue for \eqref{eq:constrainteq} of the discussion presented in \cite{Harnad:1984vk,Harnad:1985bc}\footnote{See \cite{Samtleben:2009ts,Samtleben:2010eu} for similar expansions in the case of three-dimensional supersymmetric Chern--Simons theories in the context of M2-branes.}, where it was shown that the constraint equations (in both $\CN=3$ and $\CN=4$ formalisms) of maximally supersymmetric Yang--Mills theory lead to a set of superfield equations which in turn are equivalent to the ordinary super Yang--Mills equations. This involves partially fixing gauge to reduce supergauge transformations to ordinary ones. We shall return to this issue in a forthcoming publication \cite{Saemann1}.

Moreover, it would be interesting to see if our approach could lead to a suitable description of multiple M5-branes. To this end, our equations should pass a number of consistency checks. In particular, one should develop a reasonable reduction mechanism of our equations to five-dimensional super Yang--Mills theory. The nature of this mechanism is not obvious to us at the moment. Perhaps less difficult could be the reduction of the constraint system \eqref{eq:constrainteq} to that of maximally supersymmetric Yang--Mills theory in four dimensions \cite{Harnad:1984vk,Harnad:1985bc}. Besides that, it would be important to compare the construction of non-Abelian self-dual strings with the loop space picture developed in \cite{Saemann:2010cp,Palmer:2011vx}.

\section*{Acknowledgements}
We would like to thank N.~Rink for helpful discussions. We would also like to thank the organisers of the programme `\href{http://www.newton.ac.uk/programmes/BSM/index.html}{Mathematics and Applications of Branes in String and}\linebreak\href{http://www.newton.ac.uk/programmes/BSM/index.html}{M-theory}' at the Newton Institute, Cambridge, during which this work was carried out. We are particularly grateful for assigning us the office M5. The work of CS was supported by a Career Acceleration Fellowship from the UK Engineering and Physical Sciences Research Council.

%\bibliographystyle{latexeu}
%\bibliography{littleone}

\end{document}